\documentclass[10pt,conference]{IEEEtran}
\pdfoutput=1
\usepackage{amsmath, amsfonts}
\usepackage{hyperref}
\usepackage{amssymb}
\usepackage{amscd}
\usepackage[algo2e,ruled,vlined]{algorithm2e} 
\usepackage{rotating}
\usepackage[usenames,dvipsnames]{xcolor} 
\usepackage{graphicx}
\usepackage{tikz}
\usepackage{threeparttable}
\usepackage{array, multirow}
\usepackage{color}
\usepackage{float}

\newcommand{\thm}{\begin{theorem}}
\newcommand{\ethm}{\end{theorem}}
\newcommand{\defn}{\begin{definition}}
\newcommand{\edefn}{\end{definition}}
\newcommand{\eprop}{\end{proposition}}
\newcommand{\prop}{\begin{proposition}}

\usepackage{caption}
\usepackage[caption=false,font=footnotesize]{subfig}
\IEEEoverridecommandlockouts 

\begin{document}

\title{\huge Multi-Level Anomaly Detection on Time-Varying Graph Data}

 \author{
 \IEEEauthorblockN{Robert A. Bridges\IEEEauthorrefmark{1}$^{,1}$, John Collins\IEEEauthorrefmark{1}, Erik M. Ferragut\IEEEauthorrefmark{1}, 
 Jason Laska\IEEEauthorrefmark{1} and Blair D. Sullivan\IEEEauthorrefmark{2}$^{,2}$}
 \IEEEauthorblockA{\IEEEauthorrefmark{1} Computational Science and Engineering Division, Oak Ridge National Laboratory, Oak Ridge, TN 37831 \\
 \{bridgesra,ferragutem,laskaja\}@ornl.gov, jparcoll@gmail.com}
\IEEEauthorblockA{\IEEEauthorrefmark{2} Department of Computer Science, North Carolina State University, Raleigh, NC 27695\\
 blair\_sullivan@ncsu.edu}} 
 

 \date{}

\maketitle

 {\let\thefootnote\relax\footnote{$^1$Lead author. Phone: (865) 241-0319}}
{\let\thefootnote\relax\footnote{$^2$Supported in part by DARPA GRAPHS/SPAWAR Grant
N66001-14-1-4063, the Gordon \& Betty Moore Foundation, and the National Consortium for Data Science.}}
 {\let\thefootnote\relax\footnote{This manuscript has been authored in part by UT-Battelle, LLC under Contract No. DE-AC05-00OR22725 with the U.S. Department of Energy.  The United States Government retains and the publisher, by accepting the article for publication, acknowledges that the United States Government retains a non-exclusive, paid-up, irrevocable, world-wide license to publish or reproduce the published form of this manuscript, or allow others to do so, for United States Government purposes.  The Department of Energy will provide public access to these results of federally sponsored research in accordance with the DOE Public Access Plan (\url{http://energy.gov/downloads/doe-public-access-plan}). Any opinions, findings, and conclusions or recommendations expressed in this
publication are those of the author(s) and do not necessarily reflect the views of DOE, DARPA, SSC Pacific, the Moore Foundation, or the NCDS.}}


\begin{abstract}
This work presents a novel modeling and analysis framework for graph sequences which addresses the challenge of detecting and contextualizing anomalies in labelled, streaming graph data. We introduce a generalization of the BTER model of Seshadhri et al. by adding flexibility to community structure,  and use this model to perform multi-scale graph anomaly detection. Specifically, probability models describing coarse subgraphs are built by aggregating probabilities at finer levels, and these closely related hierarchical models simultaneously detect deviations from expectation. This technique provides insight into a graph's structure and internal context that may shed light on a detected event. Additionally, this multi-scale analysis facilitates intuitive visualizations by allowing users to narrow focus from an anomalous graph to particular subgraphs or nodes causing the anomaly. 
For evaluation, two hierarchical anomaly detectors are tested against a baseline Gaussian method on a series of sampled graphs.  We demonstrate that our graph statistics-based approach outperforms both a distribution-based detector and the baseline in a labeled setting with community structure, and it accurately detects anomalies in synthetic and real-world datasets at the node, subgraph, and graph levels. To illustrate the accessibility of information made possible via this technique, the anomaly detector and an associated interactive visualization tool are tested on NCAA football data, where teams and conferences that moved within the league are identified with  perfect recall, and precision greater than 0.786.
\end{abstract}

\section{Introduction}
Social networks are playing an increasingly important role in today's society, yet extracting domain insights from their analysis and visualization remains challenging --- in large part due to their transient nature and the inherent complexity of many graph algorithms. Many social graphs naturally have (i)  labeled nodes representing individuals or entities, and (ii) an edge set that changes over time, creating a sequence or time-series of individual snapshots of the network.  A key task in understanding this data is the ability to identifying patterns and aberrations across snapshots --- specifically in a way that can pinpoint areas of interest, and provide context for results.  Unfortunately, although this time-varying labelled scenario is also natural in many other domains (e.g. cyber-security), most existing techniques for anomaly detection are either limited to static graphs or unable to ``zoom in'' on the reason a graph is identified as non-standard. The importance of context in anomaly detection is easily exemplified in a cyber-security setting, where observing an unanticipated connection (edge) between an internal IP and an external host might warrant alarm, however given the context that many similar IPs (i.e. nodes in a common community) regularly contact that host could save an unnecessary investigation. 

Here we address the problem of identifying and contextualizing anomalies at multiple levels of granularity in the sequential graph setting.  We give a novel method for anomaly detection in time-varying graph data, using hierarchically related distributions to detect related abnormalities at three increasingly fine levels of granularity (graph-, subgraph-, and node-). Probabilistic multi-scale detection relies on comparison with an underlying graph model; we use an extension (described in Section~\ref{section-gbter}) of the recent BTER model~\cite{seshadhri2012community} that enables improved prescription of community structure. To fit an instance of the model to observed graphs, we give methods for detecting communities and estimating parameters (see Section~\ref{section-fitting}). Finally, to test a newly observed graph for anomalous structure, we compute hierarchically-related probabilities from the tuned model and their associated $p$-values using a Monte-Carlo simulation. Our workflow is a streaming detection framework, where parameters are learned from previous observations, the detector is applied to new data,  then the parameters are updated to include the new graph in the observations.  

Section~\ref{section-prob-models} defines the probability calculations for two new multi-scale detectors, as well as a baseline detector similar to that of~\cite{moreno2013network} (which is limited to detecting anomalies at the graph-level).  It is important to note that performing anomaly detection using a graph's probability\textemdash as given by the model from which it was sampled\textemdash will often result in an inaccurate detector when node labels are used. This is a consequence of the likelihood of an unlabeled graph being shared by isomorphic copies distinguished by these labels and is discussed in Section \ref{section-prob-models}. We illustrate this phenomenon and provide empirical evidence that modeling a set of statistics indicative of node/subgraph interactions provides more accurate detection in two experiments described in Section~\ref{section-ed}. In Section~\ref{section-fb}, we apply our detector to NCAA Football data, establishing its accuracy in detecting variations in teams' schedules as imposed by changes in conference membership. This application naturally allows the detector to find abnormal interactions at the node (team), community (conference), and full graph (season) levels. Finally, we describe and show sample screenshots of applying our interactive anomaly visualization tool, which leverages the multi-scale analysis to enable users to easily focus their attention on the most critical changes in the data.

\section{Related Work}
In this paper, we focus on identifying anomalous instances in a sequence of graphs with common node labels.  
This problem is neither a special case nor an extension of finding anomalous parts of a single (static) graph (a much more commonly studied problem), since 
the availability of common node labels provides information not available in a single-graph or unlabeled graph ensemble problem, and new methods are required to fully exploit this information. There is limited work transforming a graph sequence to a single instance -- e.g., Eberle et al.~consider the disjoint union of subgraphs from each data instance as a single graph in~\cite{eberle2007anomaly}. For a survey on graph anomaly detection, we refer the readers to Akoglu et al.~\cite{akoglu2014graph}. 

Common techniques for finding anomalies in graphs can broadly be categorized as using compression techniques or a form of hypothesis testing with graph statistics. 
For example,  Eberle \& Holder use a compression algorithm relying on  minimum description length to detect repetitive subgraphs and identify slight deviations as anomalies~\cite{eberle2007anomaly}.  Because this technique searches for subgraphs almost isomorphic to a found normative pattern, it is a much more rigid detection framework than ours.  Hypothesis testing has a broader set of prior work, including papers of Miller et al. which use statistics based on the residual matrix~\cite{miller2013detection, miller2012goodness}. Much of this work is geared towards detecting abnormally dense communities seeded into an R-MAT graph. In \cite{miller2013detection}, the techniques are extended to accommodate the dynamic graph setting and include methods for identifying highly connected regions. Our detectors are designed to identify anomalies caused not only by abnormal density, but changes in the interactions within or between communities. A more recent hypothesis testing approach of Neville \& Moreno fits Gaussian distributions to three statistics \cite{moreno2013network}. Due to similarities with our workflow 
(using a $p$-value estimated by a Monte-Carlo simulation from a graph model to decide anomalies), we test our method against a baseline detector using similar Gaussian estimates, although we note that  \cite{moreno2013network} focused on Kronecker graphs, not the GBTER model.
Written concurrently with this work is that of Peel \& Clauset~\cite{peel2014detecting} which addresses the problem of change detection for time-varying network sequences.  Like Peel \& Clauset, we use a hierarchical generative graph model and Bayesian hypothesis testing.  Our work differs in that it introduces a new graph model (Section \ref{section-gbter}) and seeks related anomalies at different scales (as opposed to a ``shock'' that changes the overall graph structure). 

To the authors' knowledge, using multiple related detectors that respect the structure of the graph is a new technique. By design, this analysis informs an interactive tool for exploring the nature of abnormalities in each graph, a relatively unstudied aspect of graph visualization. Wong et al.~\cite{wong2008dynamic} present a multi-scale tool for exploring large graphs, informed by a clustering algorithm especially tuned to detecting star-burst patterns. In contrast, we detect dense regions as communities and integrate multi-scale visual-analytics with anomaly scores.  

This paper extends the general anomaly detection work-flow of Ferragut et al.~\cite{ferragut2012new} to hierarchically analyze graph data. The general method estimates probability models from observations and new data is declared anomalous if it has sufficiently small $p$-values.  More precisely, if a
probability distribution $P$ is estimated from observed data $x_1, ... , x_{n-1}$, the $p$-value of new data, $x_{n}$, is $p\text{-value}(x_{n}) = P(\{X: P(X)\leq P(x_n)\}).$ Notice that in the stereotypical case where $P = \mathcal{N}(0,1)$, the standard normal, the definition above corresponds to the two-sided $p$-value.
Generally, a threshold $\alpha \in [0,1]$ is set, and if $p$-value$(x_n) \leq \alpha$, $x_n$ is identified as anomalous. For streaming data, model parameters are iteratively updated to include the new observation, $x_{n}$.  Often, $\alpha$ is tuned in light of labeled results to find an acceptable balance of false vs. true positives. 
Analysis in \cite{ferragut2012new} identifies operational benefits of the method, including a theorem allowing users to regulate the expected alert rate \`{a} priori by setting $\alpha$. We utilize the framework's accommodation of any probability model in order to apply it simultaneously at hierarchical levels.  
\section{The Generalized BTER Model (GBTER)} 
\label{section-gbter}
In order to perform probabilistic anomaly detection, we need a randomized generative graph model that enables computation of probabilities for various graph configurations while accurately modeling a graph's community structure and degree sequence.  Significant prior work has  been devoted to developing such models and validating the importance of capturing both these aspects of a real-world data set (e.g.,~\cite{seshadhri2012community, barabasi1999emergence, chakrabarti2006graph, chung2002average, kolda2013scalable}). A broad survey of graph models and common graph characteristics is given in~\cite{chakrabarti2006graph}.

More specifically, motivated by social and cyber settings, we require a generative model that can accommodate observed hierarchical structure.  A natural candidate is a Stochastic Block Model, first introduced in \cite{holland1983stochastic}, which defines community membership and generates intra-community edges with an Erd\"{o}s-R\'{e}nyi (ER)~\cite{erdros1959random} model and inter-community edges with a probability that depends on the membership of their endpoints. This achieves flexible community membership and density, but the expected degree of each node is implicitly determined by the community structure and parameters. 
To improve adherence to degree distribution, one could use the Block Two-Level Erd\"{o}s-R\'{e}nyi (BTER) of Seshadhri et al.~\cite{seshadhri2012community, kolda2013scalable}, but we found the implicitly determined community structure of the model to be too limiting for matching real-world data. BTER edge generation occurs in two steps, with an ER model\footnote{denoted here as ER$(n,p)$, in which $n$ nodes are fixed and edges occur independently with probability $p$} used for intra-community edges, followed by a Chung-Lu (CL) process~\cite{chung2002average} to match a specified expected degree distribution. 

To address these challenges, we define and use a generalization of BTER that mimics its two-step edge generation process, but allows explicit prescription of the communities' size, membership, and approximate density. The remainder of this section describes this generalized version and compares it to the original BTER model. 

The generalized block two-level Erd\"os-R\'enyi (GBTER) model takes as input 
(1) the expected degree of each node, 
(2) community assignments of the nodes, i.e., a partition of the vertex set into disjoint subsets, $\{C_j\}$, and
(3) an edge probability $p_j$ for each community $C_j$.  
In the first stage of edge generation, intra-community edges are sampled from an Erd\"{o}s-R\'{e}nyi random graph model, ER($|C_j|,p_j$) for each community $C_j$. Note the expected degree of a node within $C_j$ is $p_j(|C_j|-1)$ after the first stage.  In the second stage, we define the  \textit{excess expected degree} of a node $i$, denoted $\varepsilon_i$, to be the difference between the input expected degree $\lambda_i$ and the expected degree after stage one. Formally, $\varepsilon_i : = \max( 0, \lambda_i - p_j(|C_j|-1) )$ for node $i$ in community $C_j$.  We then apply a Chung-Lu style model~\cite{chung2002average} on the  \textit{excess expected degree}-sequence, $[\varepsilon_i]_{i\in V}$.  Specifically, the probability of adding the edge $(i,j)$, is 
\begin{equation} 
\label{eqn-chunglu}
P(i,j \mid \varepsilon ) = \frac{\varepsilon_i\varepsilon_j}{\sum_k\varepsilon_k}.
\end{equation}
Note that the second stage can generate both inter- and intra-community edges. 
It is necessary that Chung-Lu inputs, $\{\varepsilon_i\},$ satisfy
 $\varepsilon_i\varepsilon_j\leq \sum_k\varepsilon_k$ for Equation \ref{eqn-chunglu} to define a probability. 
A calculation shows that the expected degree of node $i$ is indeed $d_i$ whenever 
$d_i \geq p_j(|C_j|-1)$ (i.e., the expected degree from the first-stage edges does not exceed the total expected degree of any node), and the CL model is well-defined. 

To calculate the probability of edge $(i,j)$, we condition on whether $i$ and $j$ share a community. Recall, our communities partition the set of nodes, so each $i$ is in exactly one community. If $i,j$ are assigned to the same community, $C$,
 let $p$ denote the internal edge probability of $C$, and  we see 
\begin{equation}
\label{eqn-edge-prob}
P(i,j \mid i,j \in C) = p +(1-p) \frac{\varepsilon_i\varepsilon_j}{\sum \varepsilon_k}.
\end{equation}
If $i,j$ are assigned to different communities, the edge probability is as given in Equation \ref{eqn-chunglu}. 

GBTER differs from the original BTER model by allowing greater flexibility and assignment of community membership, size, and internal edge density ($p$).  
As indicated in \cite{kolda2013scalable}, the expected clustering coefficient for an ER($n,p$) graph is $p^3$.
This implies that GBTER also allows pre-specification of each community's approximate clustering coefficient.
Note that GBTER, as used in this work, assumes node labels, but BTER on the other hand only depends on the number of nodes of each expected degree. 
This implies that edges in BTER do not occur independently (because they are conditioned on the community assignment of each node), while they are independent in GBTER. Consequently, calculating probabilities of graphs according to the BTER model is both complicated and expensive, inhibiting its use for anomaly detection.  

\section{Fitting Model Parameters}
\label{section-fitting}  
We now describe how to fit the GBTER model to a sequence of observed graphs with common node labels using 
Bayesian techniques for learning the parameters and inferring the following model inputs:
(1) the community assignments, (2) the within-community edge densities, and (3) the expected node degrees.
Once a specific instance of the model is deduced, probabilistic anomaly detectors are constructed, as detailed in Section~\ref{section-prob-models}.

In this work, a partition of the vertex set into communities is learned using the Markov Clustering (MC) algorithm \cite{van2000graph}.
We chose MC as it is known to scale well and  is easy to implement. 
To apply MC, a weighted graph is constructed from observed graphs.
Specifically, for the experiments in Section~\ref{section-ed}, the weighted graph is constructed by counting the occurrence of each edge, and for the application in Section~\ref{section-fb} exponential weights are used to down-weight older observation of edges. 
In general, any method of partitioning nodes into communities acceptable for the application at hand will suffice.
For a survey of community detection algorithms see~\cite{fortunato2010community}. 
We note that our method requires a partition of the nodes into communities but is blind to the algorithm used. 
This gives a lever for tuning between scalability and accuracy in applications. For example, communities inferred from context (e.g., grouping nodes by a known, common affiliation) can be used to obviate this step and may provide more insightful results in a real-world setting.

Given community assignments, the within-community edge densities are estimated. 
Each community, $C$, is modeled internally by an Erd\"{o}s-R\'{e}nyi random graph, ER($|C|,p$), and we seek to estimate $p$. 
Letting $k$ denoted the number of edges within the subgraph $C$, it follows that $k \sim \mbox{Binomial}(  {|C| \choose 2},p)$.
In order to use Bayesian inference, we assume $p\sim$ Beta$(\alpha,\beta)$, with prior parameters $\alpha>0,  \beta>0$, and then use the maximum posterior likelihood estimation (MPLE).  
Specifically, 
$(p\big{|}k_1, \dots, k_N) \sim \mbox{Beta}(\hat{\alpha}, \hat{\beta})$ with posterior parameters 
$$\hat{\alpha} = \alpha + \sum_i k_i,\mbox{ and }\hat{\beta} = \beta + N{|C| \choose 2} - \sum_i k_i$$
  where $k_i$ denotes the number of edges internal to $C$ observed in the $i$-th graph, $G_i$, for $i=1,\dots,N$.  
MPLE gives $p:=(\hat{\alpha} - 1)/(\hat{\alpha} + \hat{\beta} -2),$ 
the mode of the posterior.     

Lastly, the expected degree sequence must be estimated from the data. 
For a fixed node, we assume its degree, $d$, is Poisson distributed with expected degree $\lambda$, i.e. $d \sim \mbox{Poisson}(\lambda)$. 
We use the conjugate prior, $\lambda \sim \mbox{Gamma}(\alpha,\beta)$ with prior parameters $\alpha > 1,,  \beta > 1$.
This yields the posterior distribution,  
$ (\lambda \big{|} d_1, \dots, d_N) \sim \mbox{Gamma}(\hat{\alpha}, \hat{\beta})$ with posterior parameters 
$\hat{\alpha} = \alpha + \sum_id_i,$ and $\hat{\beta}  = \hat{\beta} + N,$
where $d_i$ denotes the observed degree of the node in $G_i$.  
For each node, MPLE gives its expected degree, $\lambda := (\hat{\alpha}-1)/\hat{\beta},$ 
 the mode of the posterior Gamma.
\begin{center}
\begin{figure*}
\caption{ROC Curves from Synthetic Data Experiments. Note: Gaussian Baseline only applicable to graph level detection.}
\setlength{\tabcolsep}{2pt} 
\begin{tabular}{cccc}
& \normalsize Graph Level & \normalsize Community Level & \normalsize Node Level \\
{\rotatebox[origin=c]{90}{\normalsize Experiment 1}} & \raisebox{-.5\height}{\includegraphics[width=.32\textwidth]{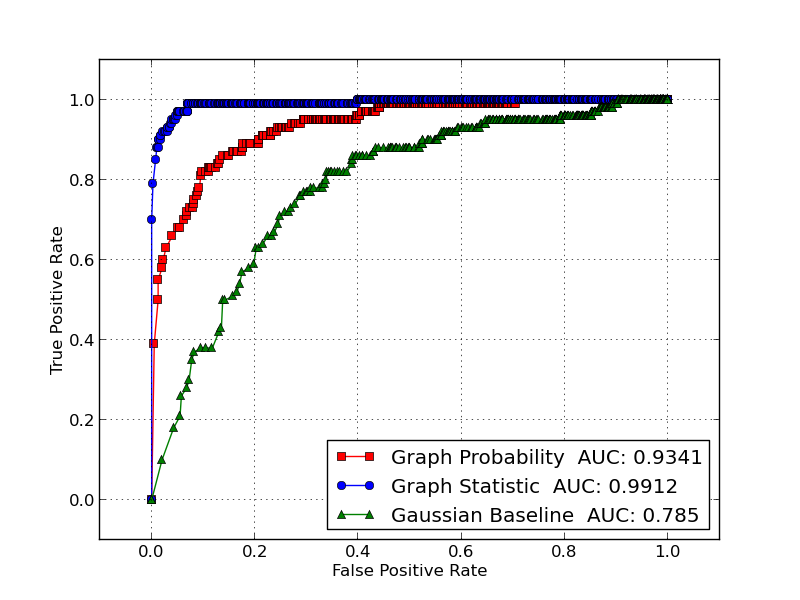} } & \raisebox{-.5\height}{\includegraphics[width=.32\textwidth]{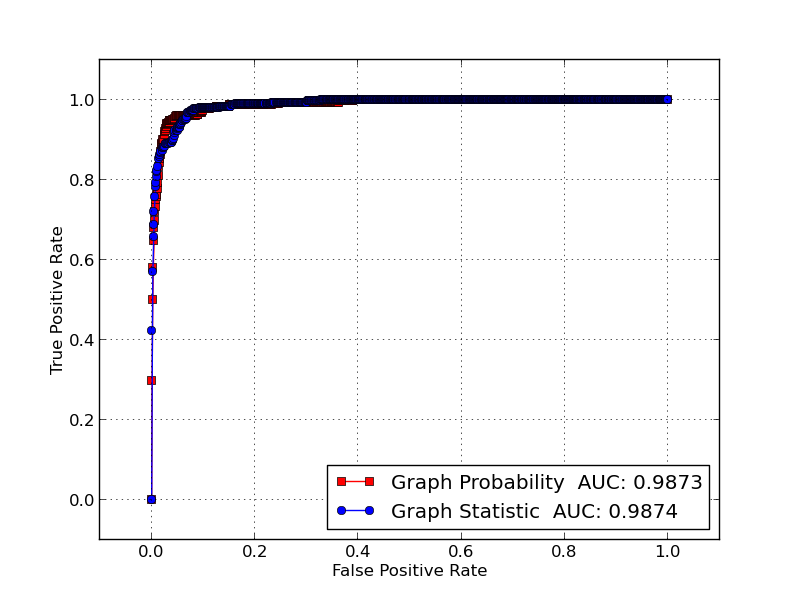}} & \raisebox{-.5\height}{\includegraphics[width=.32\textwidth]{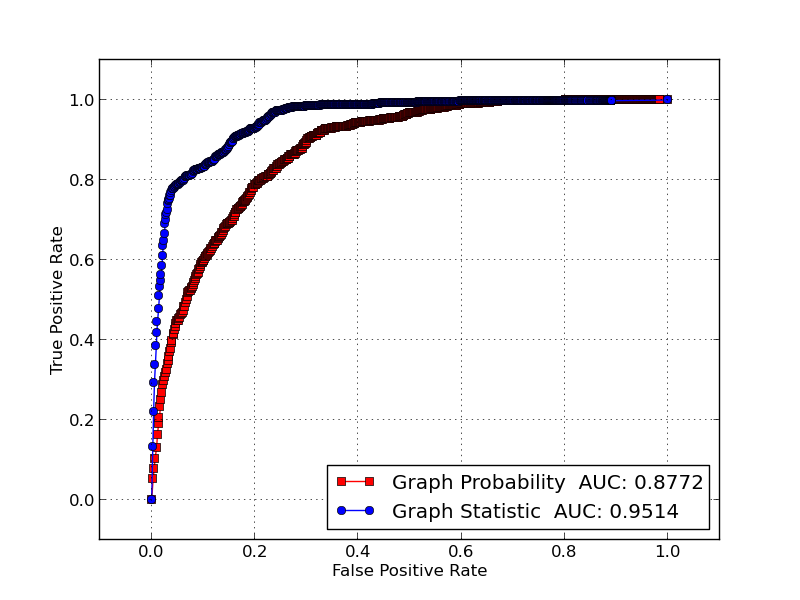} }\\
{\rotatebox[origin=c]{90}{\normalsize Experiment 2}} &  \raisebox{-.5\height}{\includegraphics[width=.32\textwidth]{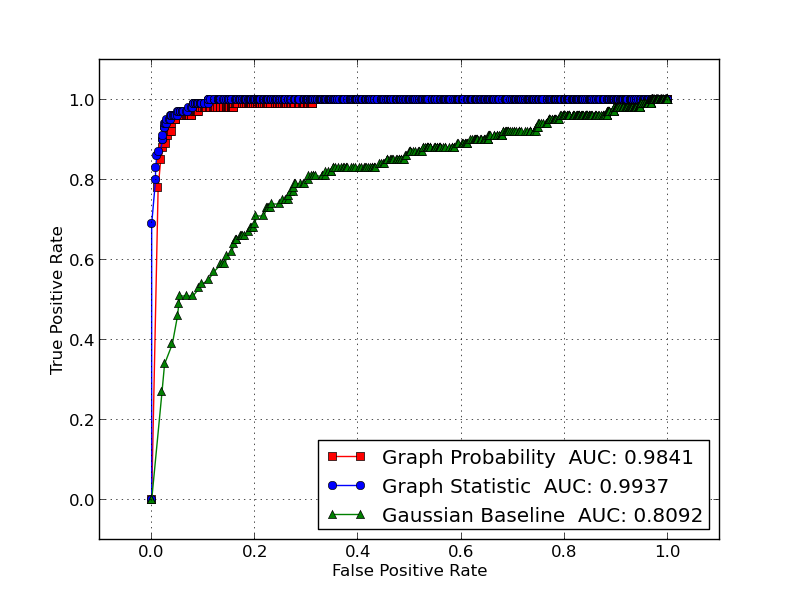} } & \raisebox{-.5\height}{\includegraphics[width=.32\textwidth]{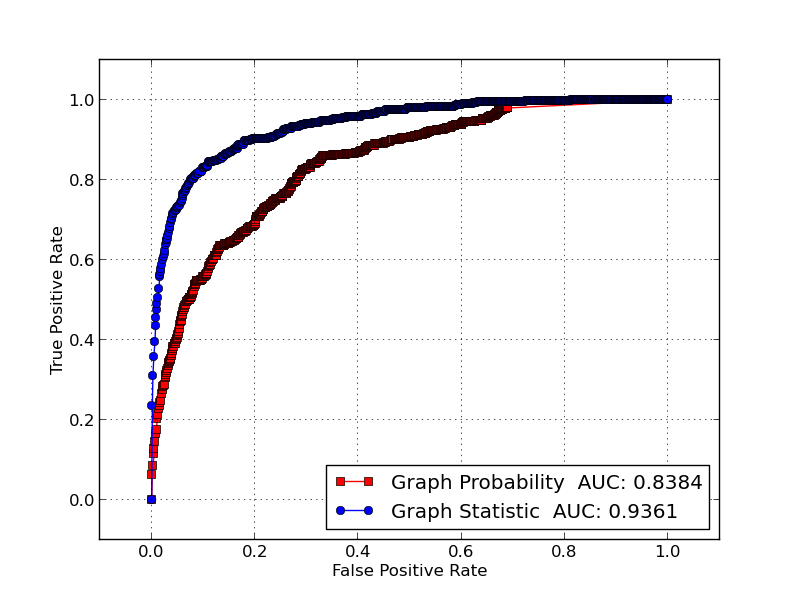}} & \raisebox{-.5\height}{\includegraphics[width=.32\textwidth]{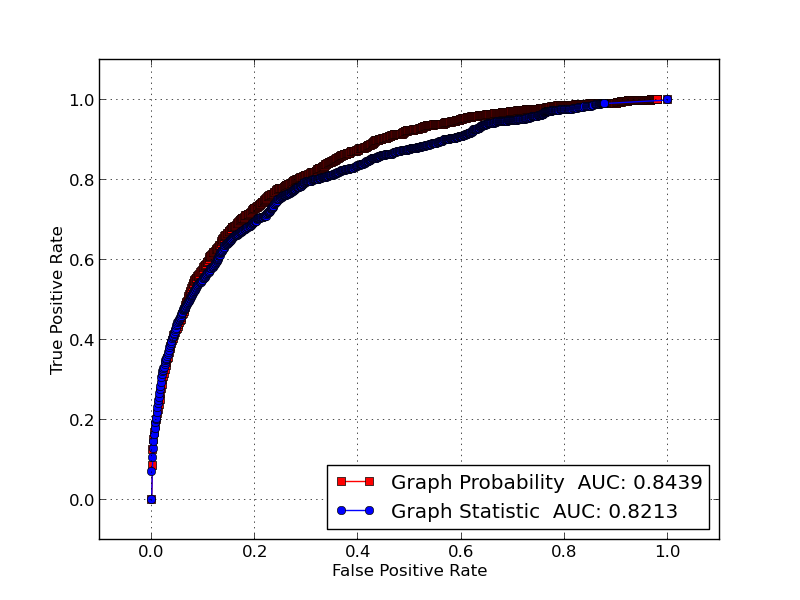} }\hphantom{aa} 
\label{fig:roc}
\end{tabular}
\end{figure*}
\end{center}
\section{Anomaly Detectors}
\label{section-prob-models}
Given an instance of a GBTER model, which defines a probability distribution on graphs, one can leverage the distribution to detect anomalies at the graph, subgraph, and node level.
This section defines two multi-scale detectors, one which uses the GBTER distribution directly, and one which leverages statistics inherent to the GBTER model. 
The Multi-Scale Probability Detector naturally uses the graph probability as determined by the GBTER model for detection, which is then decomposed into probabilities of subgraphs and nodes for hierarchical information. 
Although intuitive, this detector suffers from a few limitations, discussed below, which informs construction of the Multi-Scale Statistics Detector. 
This second detector builds from the bottom up defining the probability of a node based on the likelihood of its internal and external degree. 
Subgraph probabilities are determined by those of its member nodes, so multi-scale analysis is facilitated by both models. 
Lastly, a baseline method for detecting anomalous graphs by fitting Gaussian distributions to graph statistics, is described. We note that the Gaussian Baseline is only used for identifying anomalous graphs, as it cannot discriminate anomalies at the subgraph or node level. 
Section~\ref{section-ed} gives results of testing the three methods on synthetic (seeded) data and Section \ref{section-fb} on NCAA football data. 

\usetikzlibrary{arrows}
\usetikzlibrary{shapes}
\newcommand{\mymk}[1]{%
  \tikz[baseline=(char.base)]\node[anchor=south west, draw,rectangle, rounded corners, inner sep=2pt, minimum size=5mm,
    text height=3mm](char){\ensuremath{#1}} ;}

\begin{table*}
\normalsize
\caption{Community assignments for GBTER Experiment}
\centering
\begin{tabular}{c|cccccc}
\label{community-assign}
	& $C_1$ 	& $C_2$ 	& $C_3$ 	&$C_4$ 	& \dots & $C_{10}$\\
	\hline
$M_r$ 		& [0,1,2,3] & [4,5,6,7] & [8,9,10,11] & [12,13,14,15]  & \dots & [36,37,38,39]\\
$M_{a}$  & \mymk{[0,{\color{red}\textit{11}},2,{\color{red}\textit{4}}]} & \mymk{[{\color{red}\textit{3}},5,6,{\color{red}\textit{8}}]} & \mymk{[{\color{red}\textit{7}},9,10,{\color{red}\textit{1}}]} & [12,13,14,15]  & \dots & [36,37,38,39]\\
\end{tabular}

\begin{tablenotes}
\small
\item Note: The seeded-anomaly model $M_a$ is obtained from $M_r$ by switching the position of 2 nodes from each of the first 3 communities. Anomalous nodes shown in italicized red print, and anomalous communities are circled. 
\end{tablenotes}
\end{table*}

\subsection{Multi-Scale Probability Detector}
\label{sub-section-prob-detector}
Our first anomaly detector uses the graph probability, as given by the GBTER model, for anomaly detection. 
Specifically, given a graph $G = (V,E)$ with vertices $V$ and edges $E$, the probability of $G$ is
\begin{equation}
\label{eqn-graph-prob}
P(G) = \prod_{(i,j) \in E} P(i,j) \prod_{(i,j)\notin E}(1-P(i,j)),
\end{equation}
where $P(i,j)$ is the probability of the edge $(i,j)$ under the GBTER model, as derived in Section~\ref{section-gbter}. 
In practice, given a graph $G$, we compute it's probability using Equation~\ref{eqn-graph-prob}, then use Monte-Carlo simulation to estimate its $p$-value.  

In order to detect anomalies at different scales, the probability of a graph is decomposed into a product of subgraph probabilities. 
Specifically, we define the probability of node $i_0$ as
$$P(i_0):= \prod_{j: (i_0,j)\in E}P(i_0,j)\prod_{j:(i_0,j) \notin E}(1-P(i_0,j)).$$
It follows that $P(G) = \prod_i P(i)^{1/2}.$  
Similarly, the probability of a subgraph  $G' = (V',E')$ is $\prod P(i)^{1/2}$, with the product over $i\in V'$. 
Hence, given a partition of $V$ into communities, $\{C_i\}$, the probability of $G$ also breaks into a product of community probabilities, i.e.,  $P(G) = \prod_i P(C_i)$.   
This formulation allows anomaly detection of any fixed subgraph, in particular at the node, community, and graph level. 

The probability of sampling a graph under a given generative model is an intuitive choice for anomaly detection. 
Upon further examination, this technique will yield poor results in models where the mode of the distribution varies depending on whether or not labels are regarded. 
As an illustrative example, consider the ER model on three labeled nodes, $V = \{1,2,3\}$ with $p = 1/3$.   
The most probable unlabeled graph under this distribution has exactly one edge, and occurs with probability ${3\choose 1}(1/3)(2/3)^2= 4/9$.
Now labeling nodes, there are three different but isomorphic graphs with one edge each, namely, with edge $(1,2)$ or $(2,3)$ or $(1,3)$ only.  
But the probability of each of these one-edge graphs is $(1/3)(2/3)^2 = 4/27$, while the probability of the empty graph is $(2/3)^3 = 8/27$. 
Hence when labels are regarded, the mode of the distribution is the empty graph, not the one-edge graphs as in the unlabeled case; consequently, in this case the Multi-Scale Probability Model  will view the expected graphs as more anomalous than the less likely empty graph!
Now consider the GBTER model used in the experiment above. 
Because the probability of a within-community edge is greater than $1/2$ and inter-community edge is less than $1/2$ with the given parameters, the labeled-node mode of the distribution is the graph with every community as a clique and no other edges.  
Although this graph is unlikely to be sampled, the Multi-Scale Probability Model will regard it as the most ``normal'' possible graph. 
The conclusion of this reasoning is that using the graph's probability will produce unwarranted results, yet modeling characterizing statistics of the graph (e.g., inter- and intra-community node degrees) gives accurate detection capabilities. This is exhibited in our empirical results, and motivates the second detector.

\subsection{Multi-Scale Statistics Detector}
Our second detector is based on observing and modeling intra- and inter-community node degrees (after learning GBTER parameters).
Fix a node $i_0\in V$, and let $C$ denote $i_0$'s community, $p$ denote $C$'s intra-community edge probability, and $\lambda$ the expected degree of node $i_0$ (all as learned from fitting the GBTER model to our observations).  
We set $d_{in} := |\{(i_0,j)\in E : j\in C\}| = i_0$'s internal degree, and $d_{ex} := |\{(i_0,j)\in E : j\notin C\}| = i_0$'s external degree.
Following the ER($|C|,p$) assumption, we assume $d_{in} \sim$ Binomial($|C|-1,p$), and $d_{ex} \sim $Poisson($\varepsilon$), where $\varepsilon = \max(0,\lambda - p(|C|-1)),$ is the excess expected degree of $i_0$ (see Section~\ref{section-gbter}). 
For the Multi-Scale Statistics anomaly detector, the probability of node $i_0$ is defined as the joint probability of its degrees. 
We assume the two degrees are independent and obtain, 
\begin{align*}
P(i_0):&= P(d_{in},d_{ex}) \\
	&= {|C|-1 \choose d_{in}}p^{d_{in}}(1-p)^{|C|-1-d_{in}} \frac{e^{-\varepsilon}\varepsilon^{d_{ex}}}{d_{ex}!}
\end{align*}
Given a subgraph $G'=(V',E')$ we set $P(G') := \prod_{V'} P(i)$. 
Hence anomaly detection of any subgraph is made possible. 
 
Note that since GBTER allows both internal and external edges to be created by the second stage of the process, the model above inflates internal degree $d_{in}$ and deflates $d_{ex}$ compared to GBTER.  
Additionally, as the range of a Poisson variable is unbounded, degrees exceeding $|V|-1$ (an impossibility) are assigned positive probability by this model.
To circumvent this possibility, the truncated Poisson can be used for sampling.
In our experiments, the expected degree ($\lambda$) and expected excess degree ($\varepsilon$) are sufficiently smaller than $|V|-1$, which implies the $P(\deg(i) > |V|-1 |)$ is negligible.
Testing with and without the truncation exhibited similar results.  
	
To use either of the multi-scale detectors, we set thresholds at each level, and any node/subgraph/graph with $p$-value below the respective threshold is detected. 
The model parameters are updated upon receipt and detection of each graph. 
	
\subsection{Gaussian Baseline Detector} 
\label{section-baseline}
Our baseline method fits univariate Gaussian distributions to graph statistics and uses the product of the $p$-values for detection. From each observed graph three statistics are obtained: {average node degree} ($X_1$), {average clustering coefficient} ($X_2$), and {the spectral norm} ($X_3$). 
Calculating $X_1$ and $X_2$ from a given graph is straightforward. 
In order to calculate $X_3$, the GBTER model is used with parameters estimated as described above to produce the expected adjacency matrix $E(A)$, in which $E(A)_{i,j}$ gives the probability of an edge between nodes $i$ and $j$. 
The spectral norm is defined as the maximum modulus eigenvalue of the residual matrix $A-E(A)$. 
After computing the observed statistics, independent univariate Gaussian distributions ($\mathcal{N}(\mu_i, \sigma_i)$) are fit to each of the three statistics.  
Lastly, given a newly observed graph, $G$, with statistics $x_1, x_2, x_3$, we assign 
$$p\mbox{-value}(G) := \prod_{i=1}^3 P(X_i \leq x_i | \mathcal{N}(\mu_i,\sigma_i) ).$$
As before, $p$-values falling below a given threshold, $\alpha$, are labeled anomalous, and the three normal distributions are updated upon receipt of each new graph.  

This follows the approach of Moreno and Neville~\cite{moreno2013network}, although their work is based on Mixed Kronecker Product graph and uses average geodesic distance instead of the spectral norm we employ for $X_3$. Since the average geodesic distance is undefined for disconnected graphs, we selected  
the spectral norm based on prior use in network hypothesis testing and strong results for similar tests involving Chung-Lu random graphs~\cite{miller2012goodness}. 
While we consider this baseline a natural adaptation of \cite{moreno2013network}, the disparity in use between their and our application inhibits direct comparison. 

\section{Synthetic Graph Experiment} 
\label{section-ed} 
In order to test the anomaly detection capabilities, two hidden GBTER models are used to generate labeled data, (1)  a ``regular'' model, $M_r$, for sampling non-anomalous graphs, and (2) a  seeded-anomaly model, $M_a$, with slightly perturbed inputs to generate anomalous graphs.
To begin the experiment, 100 non-anomalous graphs are sampled from $M_r$, and the anomaly detectors are fit to the data, as described in Section~\ref{section-prob-models}.
To test the streaming anomaly detection, 500 graphs are iteratively generated and observed with every fifth graph from the seeded anomaly model.
Upon sampling a new graph, its $p-$value according to each anomaly detector is computed, and it is labeled as anomalous if it falls below a given threshold.
Similarly, the hierarchical detectors label each node and community depending on its respective $p-$value.  
Lastly, each anomaly detector's GBTER parameters are updated to include observation of the new graph. 

We conduct two experiments, both using networks of 40 nodes divided into ten equally-sized communities. For the ``regular'' model, each community is assigned a within-edge probability of $p=.8$, and the expected degrees of nodes vary in the range of five to eight according to a truncated power-law.  
To create the seeded-anomaly model for the first experiment, two nodes from each of the first three communities are  interchanged resulting in six (of 40) anomalous nodes and three (of ten) anomalous communities per anomalous graph (see Table \ref{community-assign}).
For the second experiment, community assignments are held constant, but the within-community density ($p$) of the first four communities is changed from 0.8 to 0.4 in the seeded-anomaly model, and the expected degree of the nodes in these four communities is increased by two.
This will decrease intra-community, and increase extra-community interaction for these four communities. 
All together the second experiment has four (of ten) anomalous communities, and 16 (of 40) anomalous nodes per anomalous graph. 


\newcommand\T{\rule{0pt}{2.6ex}}       
\newcommand\B{\rule[-1.2ex]{0pt}{0pt}}

\begin{table}[h!]
\caption{GBTER Experiment Results ($\alpha$ maximizing F1)}
\label{table}
\centering
\begin{tabular}{ccccc}
\hline
Method & $\alpha$ & F1 & Precision & Recall \\
\hline
\hline \T 
\textsc{\textbf{Experiment 1}} \B \\
\hline
\textbf{Graph Level}\\
\hline
Graph Probability    & 0.020 & 0.742 		& 0.678 & 0.820  \\
Graph Statistic     & 0.009 & \textbf{0.919} & \textbf{0.929} & \textbf{0.910}  \\
Gaussian Baseline    & 0.029 & 0.526 		& 0.418 & 0.710  \\
\hline
\textbf{Community Level}\\
\hline
Graph Probability & 0.019 & 0.810 		& 0.745 			& \textbf{0.887}   \\
Graph Statistic & 0.009 & \textbf{0.830} & \textbf{0.840} & 0.820 \\
\hline
\textbf{Node Level}\\
\hline
Graph Probability & 0.020 & 0.298 		& 0.239 			& 0.393\\
Graph Statistic & 0.017 & \textbf{0.547} & \textbf{0.453} &\textbf{ 0.690}\\
\hline
\hline \T
\textsc{\textbf{Experiment 2}} \B \\
\hline
\textbf{Graph Level}\\
\hline
Graph Probability    & 0.007 & 0.895 		& 0.855 			& \textbf{0.940}  \\
Graph Statistic     & 0.011 & \textbf{0.922} & \textbf{0.904} & \textbf{0.940}  \\
Gaussian Baseline    & 0.006 & 0.590 		& 0.697 & 0.510  \\
\hline
\textbf{Community Level}\\
\hline
Graph Probability & 0.062 & 0.436 & 0.390 & 0.495 \\
Graph Statistic & 0.028 & \textbf{0.654} &\textbf{ 0.620 }&\textbf{ 0.693}  \\
\hline
\textbf{Node Level}\\
\hline
Graph Probability & 0.053 &\textbf{ 0.436} & 0.368 & \textbf{0.533} \\
Graph Statistic & 0.047 & 0.434 & \textbf{0.427 }& 0.442\\
\hline
\hline
\end{tabular}
\end{table}


To evaluate the detectors' performance, the Receiver Operator Characteristic (ROC) curve, and area under the ROC curve (AUC) are displayed in Figure~\ref{fig:roc}. 
Recall that the Gaussian Baseline is only for graph level detection and thus does not contribute to the community or node level results. 
Table \ref{table} includes Precision, Recall, and F1\footnote{F1 is defined as the harmonic average of Precision, $P$, and Recall, $R$. Specifically, F1:=ave$(P^{-1}$,$R^{-1})^{-1} = 2PR/(P+R)$.} for each detector at the threshold $\alpha$ maximizing its F1 score.
In light of the ROC, AUC, Precision, Recall, and F1 scores, we see the Multi-Scale Probability Model dominates the Graph Statistic Model in most categories. 
For the full graph tests, the Gaussian Baseline is far inferior to the new models with the Multi-Scale Statistics Detector as the clear winner. 
Further, the results at all levels provide evidence that the Multi-Scale Statistics Model is the superior method, as expected after the \`{a} priori analysis given in Section \ref{sub-section-prob-detector}.


\begin{table*}[t]
\caption{Ten Most Anomalous Teams for years 2010, 2011, \& 2012 as given by Graph Statistics Detector. Threshold $\alpha = 10^{-6}$ gives perfect classification. }
\label{team-table}
\begin{tabular}{cccc||cccc||cccc}
Team & $p-$value & 2009 Conf. & 2010 Conf. & Team & $p-$value & 2010 Conf. & 2011 Conf. & Team & $p-$value & 2011 Conf. & 2012 Conf.\\
\hline
\hline
Wake Fst. & 4.84e-4 & ACC & ACC & 				Boise St. & 2.81e-18 & WAC & MWC &				Missouri & 3.95e-16 & Big-12 & SEC   \\
Wash. & 2.17e-3 & PAC-10 & PAC-10 &	 		Utah  & 1.59e-12 & MWC & PAC-10 & 				W. VA & 7.70e-14 &  Big-East & Big-12\\
San Jose St. & 2.78e-3 & WAC & WAC & 			BYU & 6.33e-10 & MWC &  -  & 					Texas A\&M & 4.03e-13 & Big-12 & SEC  \\
Utah St. & 3.05e-3 & WAC & WAC & 				Colorado & 1.70e-09 & Big-12 & PAC-10 &			Texas Chr. & 7.66e-10 & MWC & Big-12 \\ 
Tulsa & 4.32e-3 & CUSA & CUSA  &					Nebraska & 3.53e-08 & Big-12 & Big-10 & 			Temple & 1.28e-09 & MAC & Big-East  \\ 
Toledo & 1.25e-2 & MAC & MAC  & 					Wash. St. & 1.29e-06 & PAC-10 & PAC-10&		Nevada  & 2.55e-07& WAC & MWC  \\ 
San Diego St. & 1.50e-2 & MWC & MWC & 			Wash. & 2.28e-06 & PAC-10 & PAC-10  &		Fresno St. & 3.08e-07 & WAC & MWC  \\
Maryland & 2.21e-2 & ACC & ACC  &					Arizona St. & 3.28e-06 & PAC-10 & PAC-10  &		Hawaii & 4.19e-07 & WAC & MWC  \\
N.C. & 2.21e-2 & ACC & ACC &			San Jose St. & 2.26e-05 & WAC & WAC & 			Texas & 4.10e-05 & Big-12 & Big-12\\
N.C. St. & 2.63e-2 & ACC & ACC  &  	Utah St. & 2.31e-06 &  WAC & WAC & 				Miss. & 1.04e-04 & SEC & SEC
\end{tabular}
\end{table*}

\begin{table}
\centering
\caption{Ten most anomalous conferences for each year displayed with $p-$value and number of membership changes. Blue entries are \textcolor{BlueViolet}{true-positives} while red entries are \textcolor{BrickRed}{false-positives} with threshold $\alpha = 10^{-4}$.}
\label{conf-table}
\begin{tabular}{ccc||ccc||ccc}
2010 & $pv$ & $n$ & 2011 & $pv$ & $n$ & 2012 & $pv$ & $n$ \\
\hline
\hline
\textcolor{BrickRed}{ACC} & \textcolor{BrickRed}{0.000} & \textcolor{BrickRed}{0}&				\textcolor{BlueViolet}{MWC} & \textcolor{BlueViolet}{0.000} & \textcolor{BlueViolet}{3} &		\textcolor{BlueViolet}{WAC} & \textcolor{BlueViolet}{0.000} & \textcolor{BlueViolet}{5} \\
WAC & 0.001  & 0&			\textcolor{BlueViolet}{PAC-10} & \textcolor{BlueViolet}{0.000} & \textcolor{BlueViolet}{2} & 	\textcolor{BlueViolet}{Big-12} & \textcolor{BlueViolet}{0.000} & \textcolor{BlueViolet}{4} \\
CUSA & 0.090 & 0&			\textcolor{BlueViolet}{Big-12} & \textcolor{BlueViolet}{0.000} & \textcolor{BlueViolet}{2} &  	\textcolor{BlueViolet}{MWC} & \textcolor{BlueViolet}{0.000} & \textcolor{BlueViolet}{4} \\
PAC-10& 0.272 & 0 &			\textcolor{BlueViolet}{WAC} & \textcolor{BlueViolet}{0.000} & \textcolor{BlueViolet}{1} & 		\textcolor{BlueViolet}{Big-East} & \textcolor{BlueViolet}{0.000} & \textcolor{BlueViolet}{2} \\
Big-12& 0.295 & 0 &			\textcolor{BrickRed}{CUSA} & \textcolor{BrickRed}{0.000} & \textcolor{BrickRed}{0} & 		\textcolor{BlueViolet}{SEC} & \textcolor{BlueViolet}{0.000} & \textcolor{BlueViolet}{2} \\
MWC& 0.433 & 0&				SEC & 0.178 & 0 &		\textcolor{BlueViolet}{MAC} & \textcolor{BlueViolet}{0.000} & \textcolor{BlueViolet}{2} \\
Sun-Belt & 0.455 & 0 &		MAC 	& 0.211 & 0 &  		\textcolor{BlueViolet}{Sun-Belt} & \textcolor{BlueViolet}{0.000} & \textcolor{BlueViolet}{1} \\
MAC 	& 0.551 &0&				ACC & 0.287 & 0 & 		\textcolor{BrickRed}{PAC-10} & \textcolor{BrickRed}{0.000} & \textcolor{BrickRed}{0} \\
SEC & 0.639 & 0 & 			Big-East & 0.324 & 0& 	CUSA & 0.002 & 0 \\
Big-East & 0.646 & 0&		Big-10 & 0.513 & 1&		ACC & 0.965 & 0 \\
\end{tabular}
\end{table}

\section{NCAA Football Data Experiment}
\label{section-fb}
To illustrate the insight given by multi-scale anomaly detection on real-world data, the Graph Statistics Model is applied to NCAA Football data~\cite{football}. 
For comparison, we also run the Gaussian Baseline Detector.
Each season is represented as a graph with a node for each Division I team and an edge for each game played. 
Seasons 2008, 2009 are used to fit parameters of the models initially, and the streaming detection is performed on years 2010-2012. 
That is, after fitting parameters on previously observed years, the detectors give $p$-values for the newly observed season. 
 Then, the parameters are updated to include the newly observed data (and the detectors are applied to the next year). 
This dataset was chosen for two reasons, (1) NCAA conferences give a ground-truth community structure to the graph, and
(2) conference membership was relatively constant in the 2008-2010 seasons but experienced changes in 2011 and 2012.
Because teams play most of their schedule within their conference, these changes are reflected in the a season's graph and community stucture. 
Our expectation is that the 2010 graph should produce a relatively higher $p$-value (be less anomalous) than the next two years. 
Furthermore, we expect our multi-scale detector to pinpoint the conferences and teams that experienced change. 
For the experiment, the parameters are learned as discussed in Section \ref{section-prob-models}. 
Communities are detected using Markov clustering as before but with exponential down-weighting of previous years' edges as in Section \ref{section-gbter}.  
With appropriate configuration of Markov clustering parameters, the communities identified match almost identically with actual conferences, and we \`{a} posteriori label/refer to communities by the corresponding conference name for ease of discussion.

\subsection{Football Data Results}
Both the Gaussian Baseline Detector and the new Graph Statistics Detector accurately classified the full graphs (seasons), identifying 2010 as non-anomalous, and 2011 and 2012 as anomalous graphs. 
More specifically, the Gaussian Baseline reported scores of $13*10^{-5}$ for 2010, numerical 0 for 2011, and $5.2*10^{-10}$ 2012\textemdash recall this is the product of three Gaussian $p$-values attained from their CDFs.
A threshold between  $10^{-10}$ and $10^{-5}$ will give accurate classification. 
Our Graph Statistics Detector reported scores of 1.0 for 2010, and 0.0 for 2011-12, indicating that no graph sampled in the Monte Carlo simulation was more probable than the 2010 graph, and none were less probable than the 2011 (or 2012) graph. 

In addition to identifying the seasons that are/are not anomalous, our method detects the conference from the graph structure, and gives $p$-values for the conferences and individual teams that are causing the anomaly.    
Table \ref{conf-table} ranks the most anomalous conferences detected each year by the Graph Statistics Detector.  
Each conference experiencing a change in membership is detected as maximally anomalous, with $p-$value = 0.  
Across all three years, there were a total of three false-positives, and no false-negatives.
At the conference level this gives precision of  $11/14 \approxeq .786$, and perfect recall (11/11).

The results for the Graph Statistics Detector at the node level are given in Table \ref{team-table}, which details the ten most anomalous teams from each season in decreasing order along with their $p-$value and ground-truth conference memberships for the previous and current season.
We notice that a threshold of $\alpha = 10^{-6}$ gives perfect classification, identifying exactly which teams changed conferences as anomalous. 
In short, this method tells not only which graphs are anomalous, but with high accuracy can pinpoint the nodes and communities causing the anomaly.

\subsection{Interactive Data Visualization}
\begin{figure}
\centering
   \subfloat[Unprocessed 2011 Season Graph]{{\includegraphics[width=.38\textwidth]{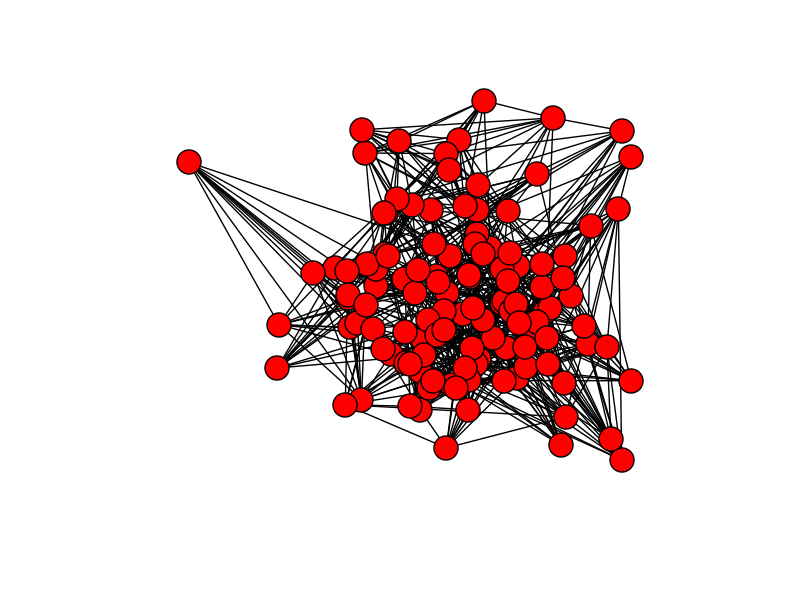} }}\\\vspace{-1.1em}
    \subfloat[Coarsened 2011 Season Graph]{{\includegraphics[width=.38\textwidth]{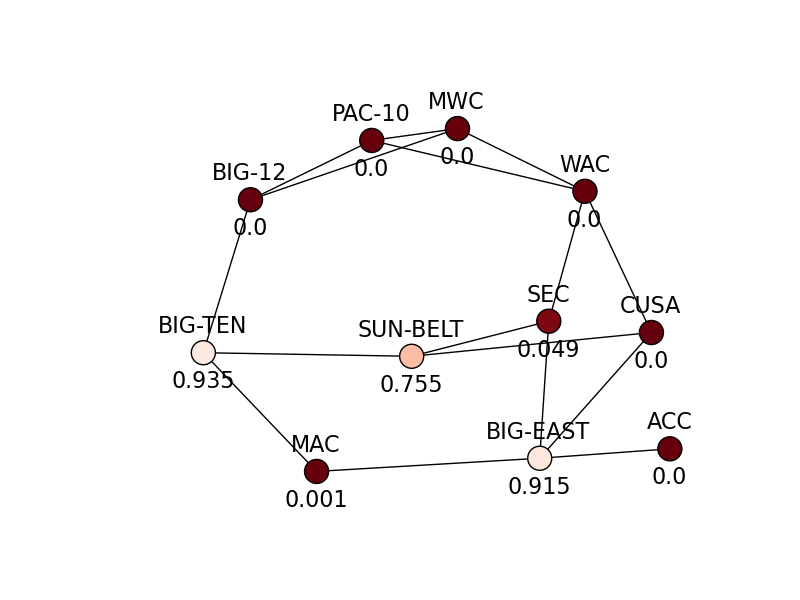} }}\\\vspace{-1.1em}
    \subfloat[2011 Mountain West Conference Graph]{{\includegraphics[width=.38\textwidth]{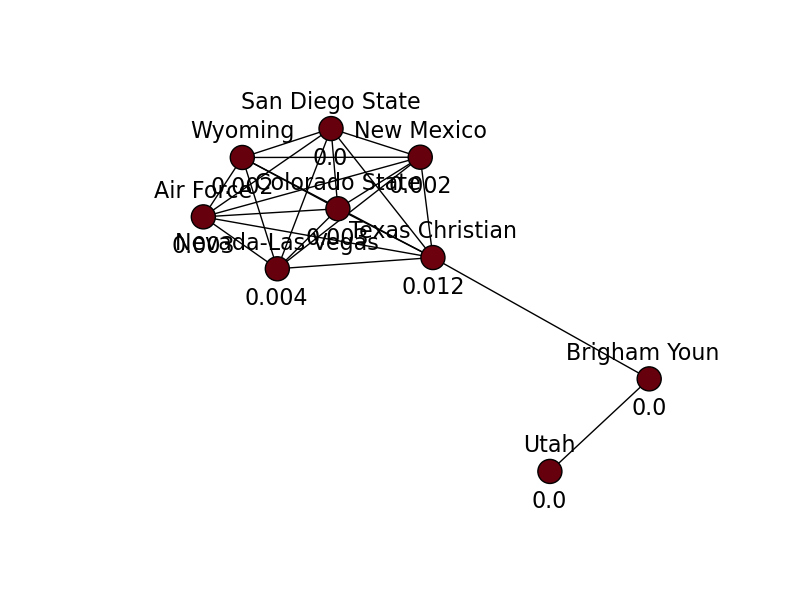} }}\\\vspace{-1.1em}
    \subfloat[2011 PAC-10 Conference Graph]{{\includegraphics[width=.38\textwidth]{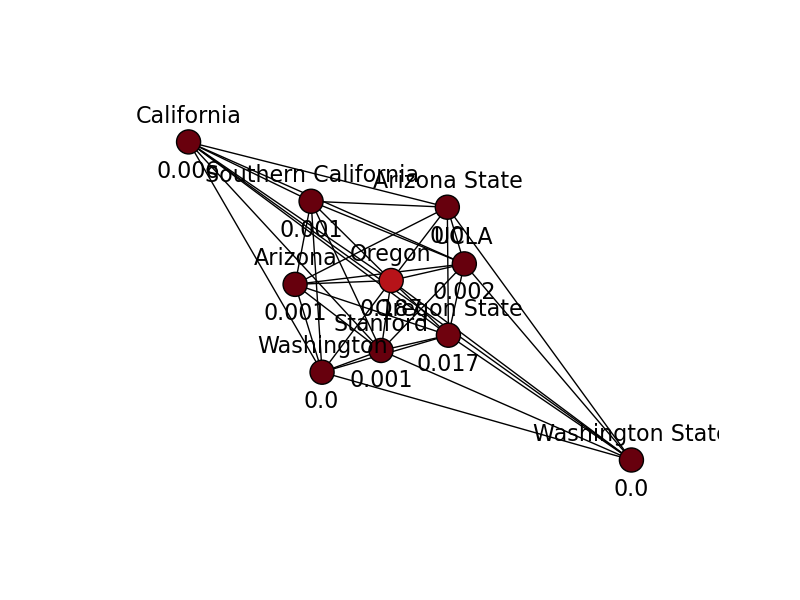} }}
         \caption{2011 Season Graph interactive visualizations screenshots (produced by MatPlotLib using NetworkX \cite{hagberg-2008-exploring} spring force directed layout) illustrate discovery of anomalies at each level. Team and conference labels were applied in post-analysis along with $p$-values in (b)-(d).  Darker colors correspond to more anomalous communities/nodes.}
	\label{fig:2011}%
\end{figure}

By design, the multi-scale detector allows users to focus attention on noteworthy communities and nodes and facilitates an interactive visualization tool for easily accessing the fine-grained structure of anomalous areas of the graph. Figure \ref{fig:2011} illustrates the benefits of this approach in screenshots from a prototype visualization.  While the 2011 graph (Figure \ref{fig:2011}.a), consists of only $\sim$130 nodes and has well-defined community structure, an unprocessed visualization provides little insight into the anomalous sections of the graph. Alternatively, coarsening and displaying only ``super''-nodes representing communities and using darker shades to indicated increased anomalousness, obviates the communities of interest (Figure \ref{fig:2011}.b). 
In addition, our tool allows conference names and $p$-values to be automatically displayed so contextual information from the analysis and the domain are easily absorbed by a user. Conference nodes are clickable, and selection displays the inter-conference subgraph, again with nodes shaded to indicate anomalousness of the teams they represent. This setup facilitates interactive exploration of anomalies, and the contexts in which they occur.
For example, clicking on the Mountain West Conference (MWC) node displays the graph in Figure \ref{fig:2011}.c, from which it is immediately apparent that, while Utah and Brigham Young were previously members of that community, they cease to participate in the MWC. The PAC-10 Conference subgraph (Figure \ref{fig:2011}.d) exhibits high density, but each node is very anomalous.  This indicates that the interaction outside the conference has changed and referencing the tables confirms that new teams, namely Utah and Colorado, are now in this conference. 
Altogether, the framework for multi-scale detection yields analytic results that are readily input into an interactive visualization. 
Upon detection of an anomalous graph, users can now zoom into areas of interest, and form and resolve hypotheses about how the anomaly occurred. 

\section{Conclusions and Future Work}
This work addresses the challenge of identifying anomalies in a sequence of graphs with emphasis on facilitating an understanding why and how a particular graph deviates from normal to put he anomalies in context. We developed and tested a framework for identifying anomalies in a time-series of graphs at three hierarchical levels of granularity.  We introduce GBTER, a generalization of the BTER generative graph model, that allows more accurate prescription and modeling of community structure, and build two hierarchical, streaming anomaly detectors, one based on the graphs probability and the other on statistics describing node and community interactions. 
Our \`{a} priori analysis predicts  the statistics-based detector will produce greater accuracy, and this is confirmed in tests on synthetic data where ground-truth is known at the node, subgraph, and graph levels. Additionally, both detectors outperform a baseline detector that fits Gaussian distributions to observed statistics of the full graph. 
In order to illustrate the insight facilitated by the multi-scale detection capability, the superior multi-scale detector is applied to NCAA football data. In both the synthetic experiment and the application to NCAA data, the Multi-Scale Statistics Detector was able to accurately pinpoint anomalies at the node, subgraph, and graph level, exhibiting the advantage of drilling into anomalous graphs to see exactly what has deviated from expectation. 
A preliminary visualization informed by the analytics is given for this example. 
We believe applying this method to other time-sampled social networks will enable discovery of underlying structure as well as anomalies and the context in which they occur. 

While investigations of scalability are outside the scope of this work, we expect applications of this approach to necessitate larger data, and we address the bottlenecks in the current implementation. Firstly, this approach requires a partition of the nodes into communities, but is agnostic to the method used. Hence, we have the ability to optimize performance by the partitioning algorithm chosen.  
As mentioned above, using communities known from context (e.g., assuming knowledge of the NCAA conferences each year) can obviate this step and provide groupings that are familiar to the operator. 
Secondly, estimating the $p$-values of a given distribution can be computationally expensive, especially if it requires sampling large graphs and calculating their probabilities. 
In general, importance sampling, in which one over-samples from a subset of the event space, can aid in Monte-Carlo simulations, although further research is required to optimize performance gains for our needs.  
Thirdly, the choice of probability models of the parameters could be changed to admit easier $p$-value computation. For example, multinomials become robust with abundant observations. In a specific application, flexibility in the modeling may yield increased performance with negligible effects on accuracy.  
Lastly, adapting the overall workflow to fit a specific application may admit performance gains. For example, updating parameters less often (in a batch process periodically) or discarding anomalous data from the update observations are options that have yet to be explored. 
In summary, while the current implementation is suitable only for small datasets, the approach has promising scalability and should be adaptable to high-volume and/or large-network settings.

\bibliographystyle{IEEEtran}
\bibliography{IEEEabrv,sdm}

\end{document}